\documentclass[11pt]{article}
\usepackage[margin=2cm]{geometry}
\usepackage{comment}
\usepackage{amsmath}
\usepackage{amsthm}
\usepackage{amssymb}
\usepackage{graphics}
\usepackage{graphicx}
\usepackage{latexsym}
\usepackage{indentfirst}
\usepackage{graphicx}
\usepackage{amsmath}
\usepackage{amssymb}
\usepackage{amsthm}
\usepackage{bm}
\usepackage{relsize}
\usepackage{titling}
\usepackage{titlesec}
\usepackage{tikz}
\usepackage{float}
\usepackage{subcaption}
\usetikzlibrary{arrows}
\usepackage{hyperref}
\newtheorem{theorem}{Theorem}

\newtheorem{proposition}[theorem]{Proposition}

\begin{document}

\title{\rule{\textwidth}{2pt}
\textbf{A regime switch analysis on Covid-19 in Romania}
\rule{\textwidth}{2pt}
}
\author{
\textbf{Marian Petrica}\\
{\small Faculty of Mathematics and Computer Science}\\
{\small University of Bucharest}\\
{\small Institute of Mathematical Statistics and}\\
{\small Applied Mathematics of the Romanian Academy}\\
{\small marianpetrica11@gmail.com}
\and
\textbf{Radu D. Stochitoiu}\\
{\small Faculty of Automatic Control and Computers}\\
{\small Polytechnic University of Bucharest}\\
{\small radu.stochitoiu@gmail.com}
\and
\textbf{Marius Leordeanu}\\
{\small Institute of Mathematics of the Romanian Academy} \\
{\small Polytehnic University of Bucharest}\\
{\small leordeanu@gmail.com}
\and
\textbf{Ionel Popescu (corresponding author)}\\
{\small Faculty of Mathematics and Computer Science}\\
{\small University of Bucharest}\\
{\small Institute of Mathematics of the Romanian Academy}\\
{\small ioionel@gmail.com}\\
}
\date{\vspace{-1ex}}
\maketitle

\begin{abstract}  In this paper we propose a three stages analysis of the evolution of Covid19 in Romania.  

There are two main issues when it comes to pandemic prediction.  The first one is the fact that the numbers reported of infected and recovered are unreliable, however the number of deaths is more accurate.  The second issue is that there were many factors which affected the evolution of the pandemic.  

In this paper we propose an analysis in three stages.  The first stage is based on the classical SIR model which we do using a neural network.  This provides a first set of daily parameters.   

In the second stage we propose a refinement of the SIR model in which we separate the deceased into a distinct category.  By using the first estimate and a grid search, we give a daily estimation of the parameters.  

The third stage is used to define a notion of turning points (local extremes) for the parameters.  We call a regime the time between these points.   

We outline a general way based on time varying parameters of SIRD to make predictions.  

\end{abstract}

\section{Introduction}

In 2019, Covid19, a virus from the coronavirus family appeared and spread around the world very quickly.  This changed dramatically our world as we knew it. 


On 31$^{st}$ of December 2019, the first cases of infection with an unknown virus causing symptoms similar to those of pneumonia were reported in China, to the World Health Organization.  


Within less than 3 months COVID-19 outbreak has become a global pandemic, spreading across almost all countries all over the world. 

The fast-evolving spread of the new coronavirus, which has been officially declared a pandemic, is represented below. The charts in Figure \ref{f:01} show the countries where there have been reported at least 1000 cases of COVID-19 infection, at the mentioned date:

\begin{figure}[H]
  \centering
  \begin{subfigure}[b]{0.4\linewidth}
   \includegraphics[width=\linewidth]{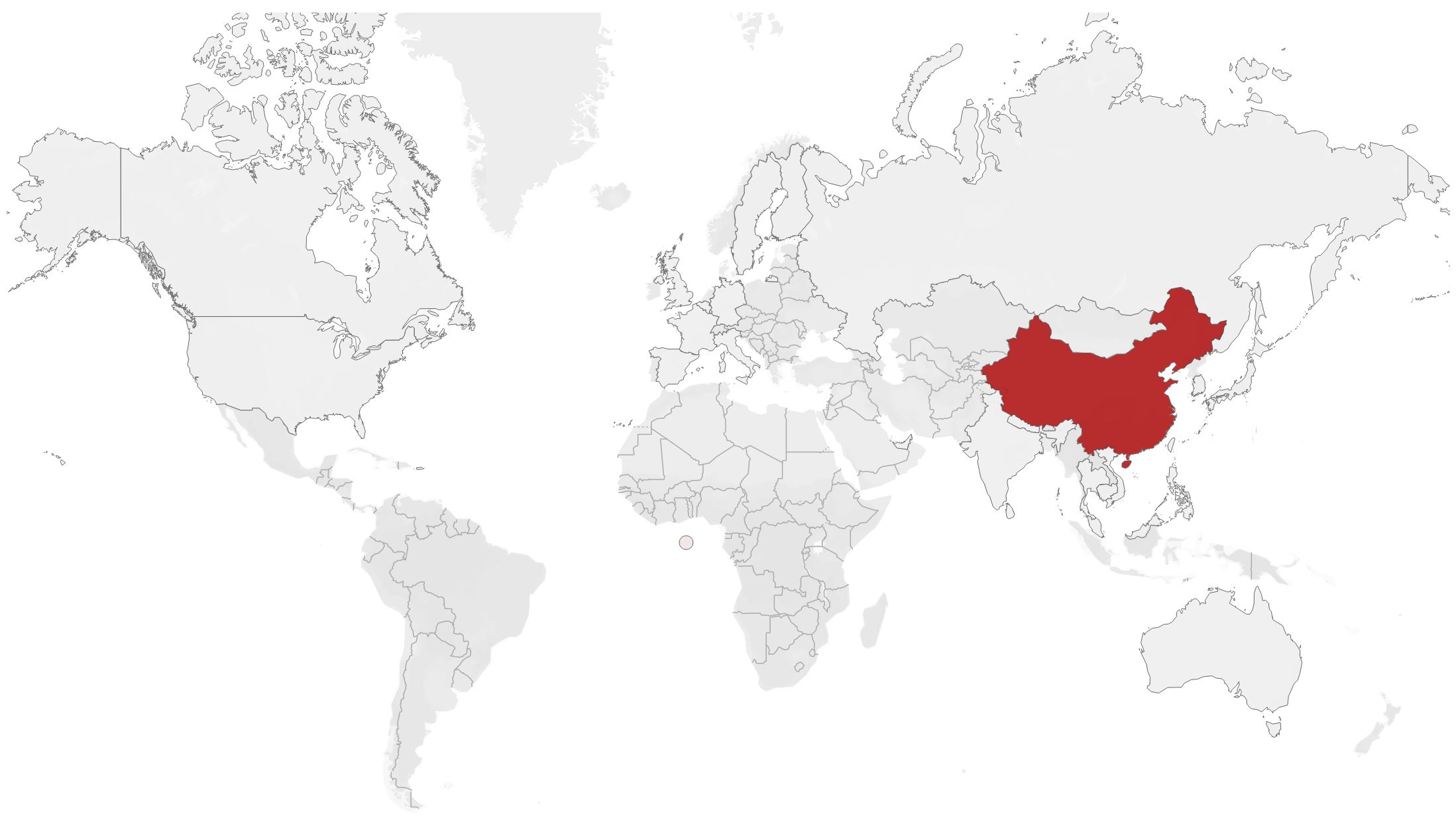}
   \caption{9$^{th}$ of February 2020.}
  \end{subfigure}
  \begin{subfigure}[b]{0.4\linewidth}
   \includegraphics[width=\linewidth]{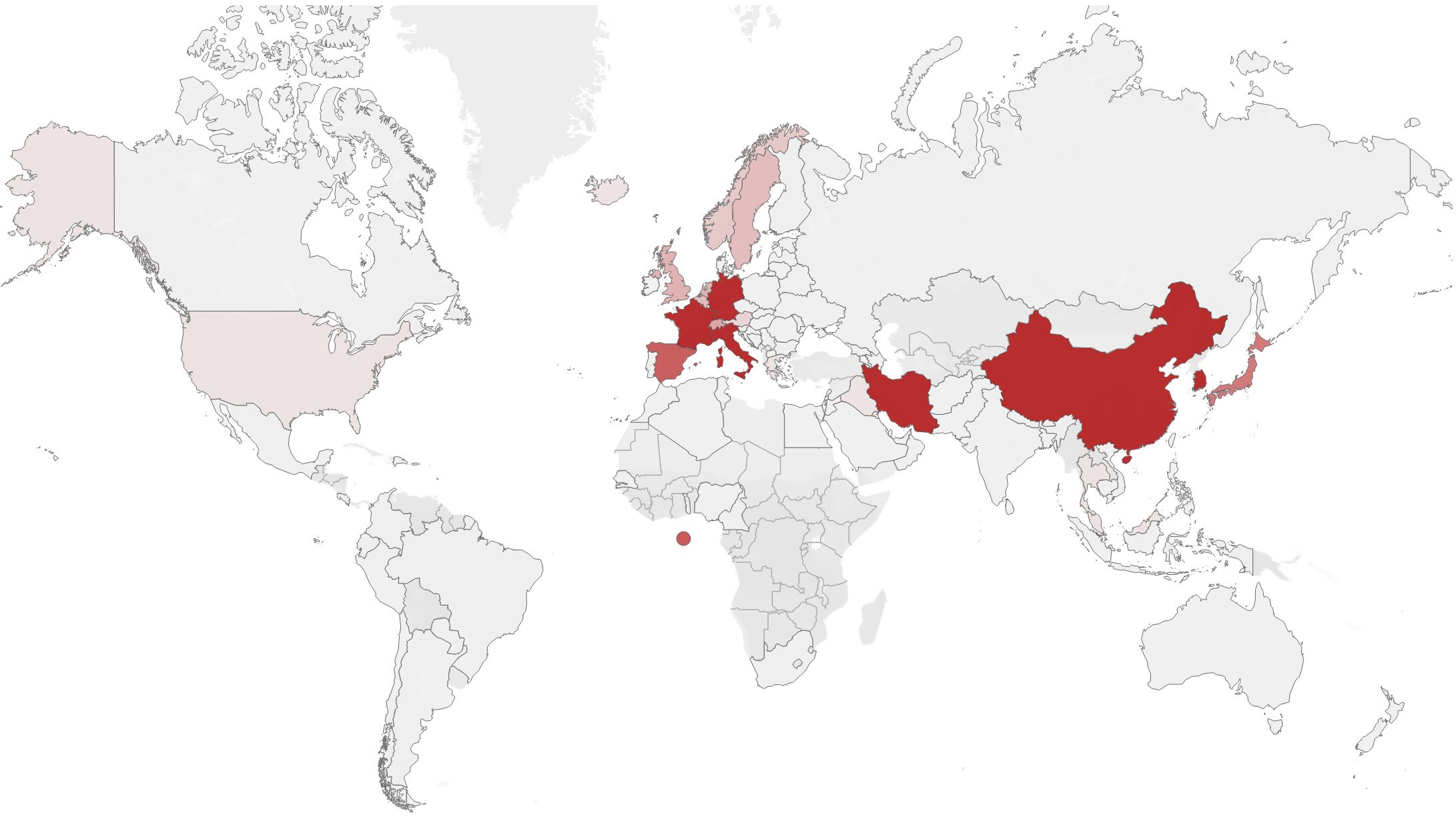}
   \caption{8$^{th}$ of March 2020.}
  \end{subfigure}
  \begin{subfigure}[b]{0.4\linewidth}
   \includegraphics[width=\linewidth]{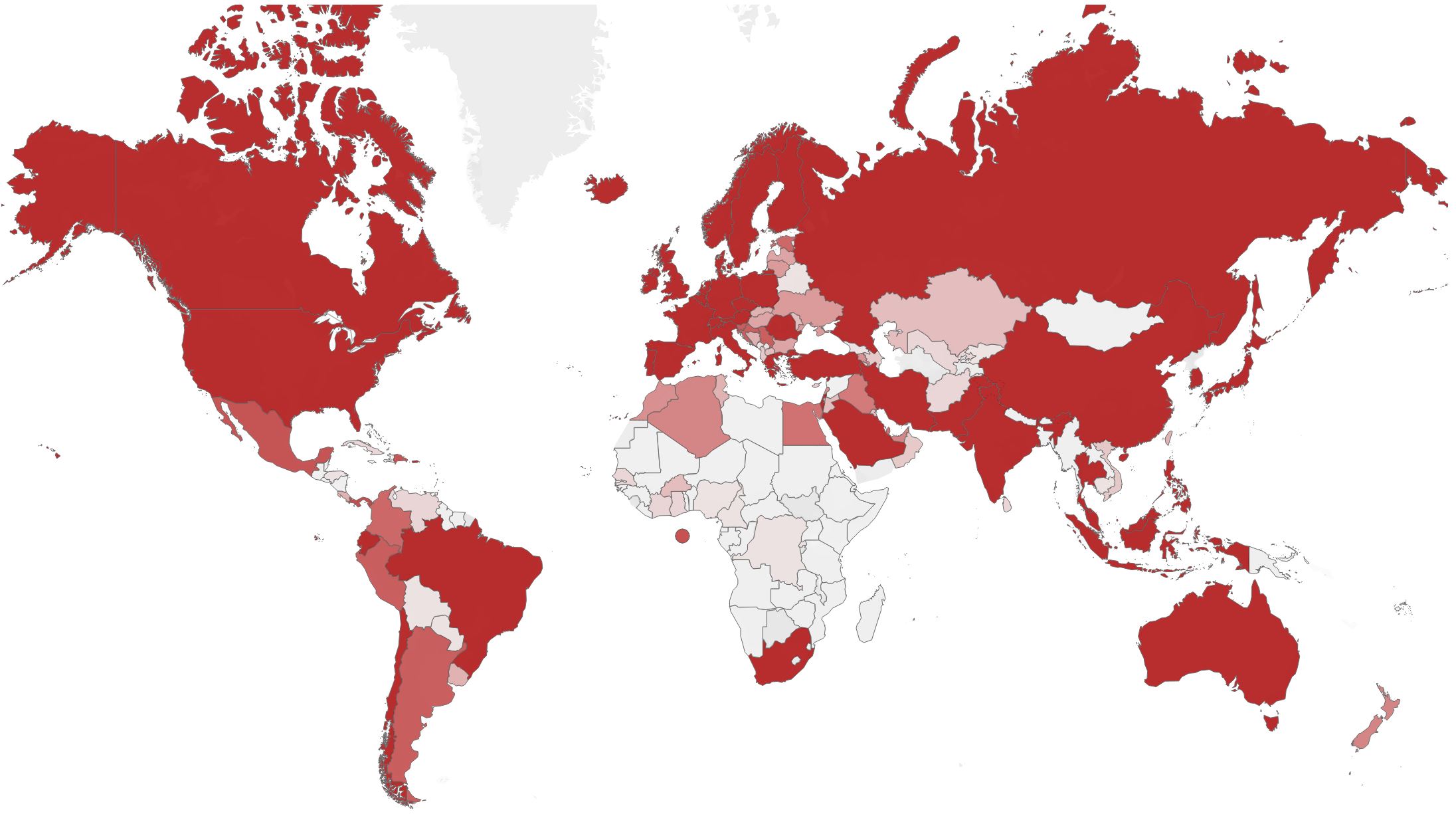}
    \caption{29$^{th}$ of March 2020.}
  \end{subfigure}
  \caption{ The spread of the Covid19 around the world during the first part of 2020.  As one can see, it affected the whole world in a very short time.  This map was generated with Tableau Desktop Software, version 10.4, \url{www.tableau.com}  }
  \label{f:01}
\end{figure}

While at the beginning of February 2020 the virus was still affecting mainly China, it has started to spread rapidly to other countries, 
and by the end of March 2020, the outbreak was present on all continents, affecting most of the countries in the world, which led the World Health Organization to officially name it a pandemic.

\subsection{The main ideas of this paper}


A basic tool in analyzing the spread of the virus is the mathematical modeling.  There is a growing body of mathematical models used at the moment as for a small sample by no means exhaustive see \cite{wakefield2019spatio,choisy2007mathematical, grassly2008mathematical, kucharski2020early,sardar2020assessment,schuttler2020covid,ferguson2020report,tsay2020modeling}.  

One of the characteristics of this pandemic is that there were many changes in the evolution.  On one hand, the political decisions changed the course of the spread at the beginning with various measures, like quarantine, isolation, work from home and so on.  Later on, other measures like relaxation, mask mandates, summer versus winter times, school openings and closing, election times, vaccination campaign, new variants of the virus and the travel ban lifting led to many changes in the status of the pandemic. 

Any reasonable parametric model of the epidemics has a major difficulty, namely the assumption that the parameters are constant over time is unrealistic.  However, what one can still do is to use the models on short periods of time and then reassemble the local behavior to get a more general picture.  This is our guiding principle in this paper.  

Our approach to modelling the Covid19 evolution in Romania is in three stages.  

The starting point is the standard SIR model initiated in  \cite{rossjohn} and later investigated in depth in \cite{kermack1991contributions1,kermack1991contributions2,kermack1933contributions3}. The basic SIR model uses the two basic parameters, the infection rate $\beta$ and the recovery rate $\gamma$.  They are assumed constant over time, however as we pointed out, the parameters of the model vary over time. However, we exploit this model on short periods of time where the assumption of constancy of parameters is still reasonable and thus we get daily estimates of the parameters based on $14$ days of data.  

We should point out that 14 days seems to be a natural choice in the analysis for many reasons.  For instance, the relative average period of recovery is 14 days.  On the other hand, the infection takes some time to fully manifest.  Also the global impact of a new variant of the virus takes a number of weeks till it is observed at large scale.  Particularly to Romania, the health units have the obligation to report the cases involving Covid19 with an accepted delay of 14 days.    

The mathematical basis for our first stage of the estimation is Proposition~\ref{p:1} which states that given a SIR model with constant parameters and data for two distinct days, we can completely determine the parameters.  We exploit this results in combination with a neural network to do the estimation inspired by \cite{param_estim2012,param_estim20202,param_estim2020_conf}.  More details are outlined in Section~\ref{s:stage1}.  We only point out that the construction of this neural network is driven by the SIR model alone. We generate data and then train a neural network to learn the parameters.   Using then the data, namely the number of infected and recovered, we estimate the main two parameters $\beta$ and $\gamma$ of the SIR model for short periods of time.  This neural network construction could potentially be used in a more general framework of dynamical systems.  

The second stage is driven by the idea that the number of deaths is more accurate and more reliable.   Thus we propose a change of the SIR model to account for the dead as a separate category and create a differential equation associated to deaths. The idea is that the infected people are evolving into either recovered, as in the standard SIR model, or die. The parameters are now, $\beta$, the infection rate, $\gamma_1$, the recovery rate and $\gamma_2$ the death rate, which models the rate at which the infected pass away. 

The basic idea here is to take the outcome of the SIR estimates of the parameters $\beta$ and $\gamma$ as a first round of approximations and then proceed to a grid search of $\beta, \gamma_1,\gamma_2$ around the suggested values from the first stage so that the model matches the observed number of deaths.  We do this again, daily, utilizing the previous 14 days to have a more realistic estimates of the parameters.  

The third stage is the definition of a regime. We look at the parameters $\beta,\gamma_1,\gamma_2$ and identify the \emph{turning points} (maximum/minimum) of the parameters.  We pay more attention to $\gamma_2$ as predicting the number of deaths is presumably more important for the preparation of the medical units involved with fighting the pandemic. Though we payed more attention to $\gamma_2$, the other parameters have extreme points approximately around the same values.    

Having completed the three stages, we can actually use the analysis for predictions.  We outline this by using a time varying SIRD model and natural estimates of the parameters using regression lines constructed in terms of 7 previous days.   Based on this we show how one can make predictions.  It turns out that for the prediction of deaths, this works well with two weeks of prediction and still reasonably well for three weeks forward.  The predictions cease in the proximity of the turning point justifying again the nomenclature of turning points.  
We insist on the methodology of the approach and complement this with numerical calculations and it can be extended to evolution equations for other diseases.  

The organization of the paper is as follows.  In Section~\ref{s:data} we show the anomalies in data and then how we cleaned and adjusted it. 

The main method is outlined in Section~\ref{s:method}.  This is composed of Section~\ref{s:stage1} where we introduce the SIR model.  We provide here the main mathematical result, namely Proposition~\ref{p:1} whose proof we postponed in the Appendix.  Next, in Section~\ref{s:2} and Section~\ref{s:ne} we present the construction of the neural network and the first estimates of the parameters, that we get using the neural network. We continue then with Section~\ref{s:stage2} where we introduce the SIRD model and in Section~\ref{s:sirdest} we show the numerical scheme for the estimations.  Furthermore in Section~\ref{s:stage3} we provide the definitions of turning point and the regime, which is completed by the  introduction of time varying SIRD model.  

In Section~\ref{s:pred} we provide an approach for the predictions using the estimates already done and we illustrate this with the case of prediction of deaths.   

The last part, Section~\ref{s:6},  is for concluding remarks.  Finally, the Appendix~\ref{t:2} provides the proof of Proposition~\ref{p:1}.

\section{The Data} \label{s:data} 

We import the data from \url{https://datelazi.ro} which keeps a record of all the data during the pandemic in Romania starting with 17th of March 2020. We limit the data till 1st of February 2022.  

One of the issues is that a first look at the raw data reveals an extreme spike in the number of recovered.  This is due to the fact that the definition of recovered patient changed in October 2020.  This added in one single day a very large number of recovered, approximately 44000 cases, which is the same to the cumulative  number of cases till that day. We redistributed these extra cases proportionally to the previous days.   

\begin{figure}[H]
    \includegraphics[width=.49\textwidth]{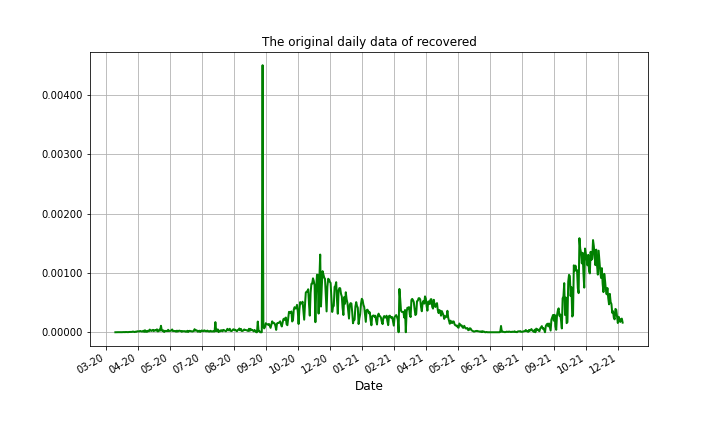}\hfill
    \includegraphics[width=.49\textwidth]{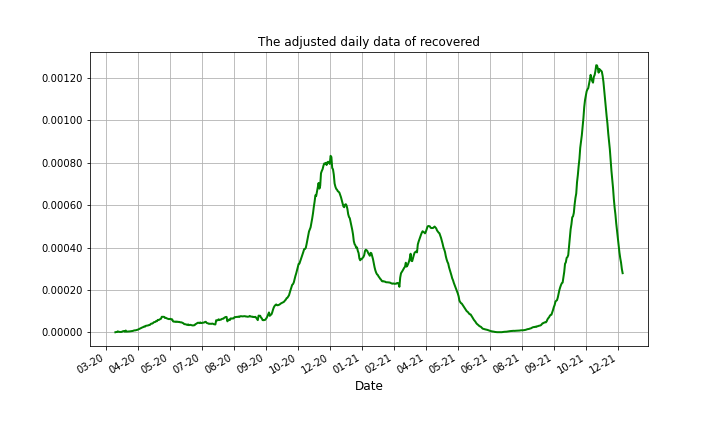}
    \\[\smallskipamount]
    \includegraphics[width=.49\textwidth]{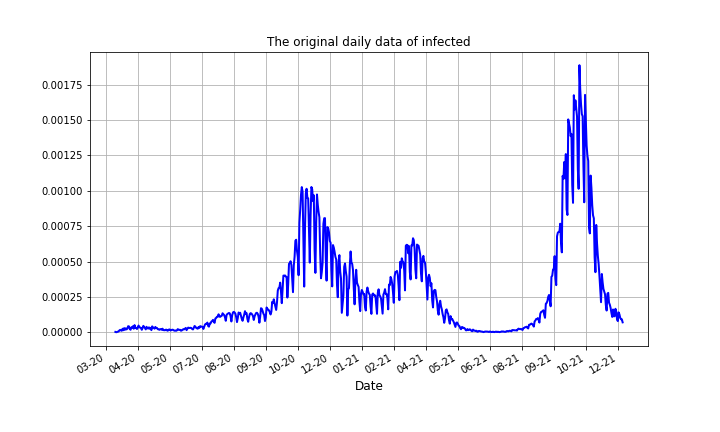}\hfill
    \includegraphics[width=.49\textwidth]{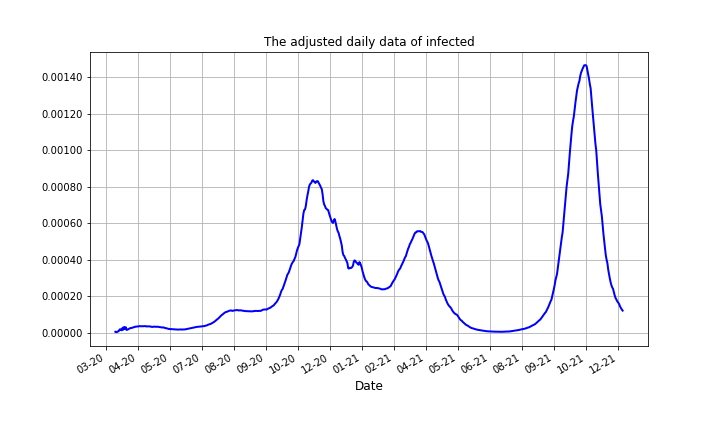}
    \\[\smallskipamount]
    \includegraphics[width=.49\textwidth]{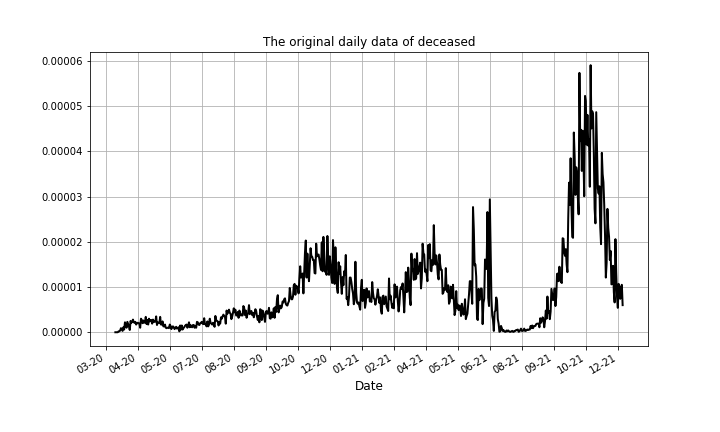}\hfill
    \includegraphics[width=.49\textwidth]{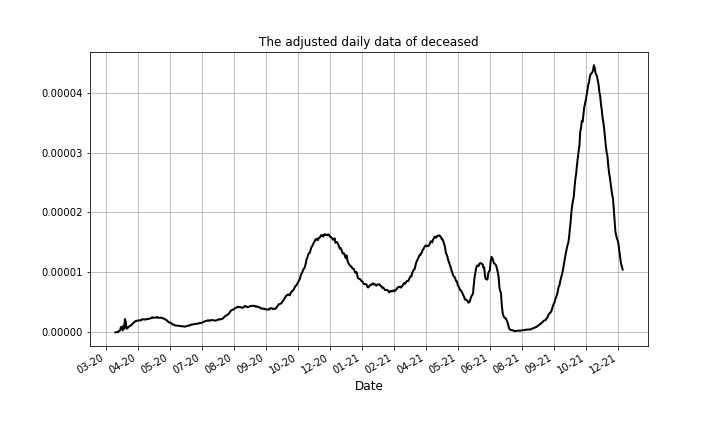}
    \caption{Data adjustments according to the actual practice of reporting the Covid19 numbers in Romania.  The rows (from top to bottom) present the number of recovered, infected and dead individuals.  The left column represents the raw data and the right column represents the adjusted data as we presented.  As a detail, notice the scale and the spike in the first picture which is adjusted as we pointed out.  The data is scaled by $10,000,000$.}
    \label{i:2}
\end{figure}

The second correction is due to the fact that there are periods of zero reported numbers.  This can range from a few days to almost three weeks.  In addition, the reports during the weekends differ substantially from the reports during the weekdays.  Also, by law, the medical centers have a flexibility of reporting the data with a delay of up to two weeks.  Therefore, in order to alleviate these irregularities we use a moving average of two weeks.  
The results are presented in Figure~\ref{i:2} with the data before and after the cleaning.  

\section{Methodology}\label{s:method}

In this section we outline the principles which guided us in this paper.  

There are three main stages.  The first stage consists in fitting a neural network on a SIR model to get a first round of parameters.  Due to the uncertainties in the number of infected and recovered, we can not fully trust these results.  

The second stage is to fit the parameters based on the more reliable data, namely the number of deceased people.  To achieve this we propose a SIRD model which combined with the grid search provides a final estimate of the parameters.

The third and final stage introduces the definitions of turning point and regime, notions that we use in order to make predictions.


\subsection{Stage I - SIR model and the first round of daily estimates}\label{s:stage1}

\subsubsection{The SIR Model}\label{s:1}

The first attempts of developing a mathematical model of the infectious diseases spreading were made at the beginning of the twentieth century. One of the most important models that can describe infectious diseases is the SIR model. The first ones that developed SIR epidemic models were Bernoulli, Ross, Kermack-McKendrick and Macdonald.

The SIR model is a mathematical model that can be used in epidemiology in order to analyze, at a given time for a specific population,  the interactions and dependencies between the number of individuals who are susceptible to get an infectious disease, the number of people who are currently infected and those who have already been recovered or have died as cause of the infection. This model can be used to describe diseases that can be contracted just one time, meaning that a susceptible individual gets a disease by contracting an infectious agent, which is afterwards removed (death or recovery).

It is assumed that an individual can be in either one of the following three states: susceptible ($\bar{S}$), infected ($\bar{I}$) and removed/ recovered ($\bar{R}$). This can be represented in the following mathematical schema:

\tikzstyle{int}=[draw, fill=blue!20, minimum size=2em]
\tikzstyle{init} = [pin edge={to-,thin,black}]
\begin{center}
\begin{tikzpicture}[node distance=3 cm,auto,>=latex']
    \node [int] (a) {Susceptible};
    \node [int] (c) [right of=a] {Infected};
    \node [int] (d) [right of=c] {Removed};
    \node [coordinate] (end) [right of=c, node distance=2cm]{};
    \path[->] (a) edge node {$\bar{\beta}$} (c);
    \draw[->] (c) edge node {$\gamma$} (end) ;
\end{tikzpicture}
\end{center}

where:
\begin{itemize}
\item $\bar{\beta}$ = infection rate
\item $\gamma$ = removed rate.
\end{itemize}

We consider $N$ as the total population in the affected area. We assume $N$ to be fixed, with no births or deaths by other causes, for a given period of n days. Therefore, $N$ is the sum of the three categories previously defined: the number of susceptible people, the ones infected, and the ones removed:
\[
N = \bar{S} + \bar{I} + \bar{R}.
\]

Therefore, we analyze the following SIR model: at time $t$, we consider $\bar{S}(t)$ as the number of susceptible individuals, $\bar{I}(t)$ as the number of infected individuals, and $\bar{R}(t)$ as the number of removed/recovered individuals. The equations of the SIR model are the following:

\begin{equation}\label{eq0}
\begin{cases}
\frac{d \bar{S}(t)}{dt}=-\frac{\bar{\beta} \bar{S}(t) \bar{I}(t)}{N} \\
\frac{d\bar{I}(t)}{dt}=\frac{\bar{\beta} \bar{S}(t) \bar{I}(t)}{N}-\gamma \bar{I}(t) \\
\frac{d\bar{R}(t)}{dt}=\gamma \bar{I}(t)
\end{cases}
\end{equation}
where:
\begin{itemize}
 \item  $\frac{d \bar{S}}{dt}$ is the rate of change of the number of individuals susceptible to the infection over time;
\item $\frac{d\bar{I}}{dt}$ is the rate of change of the number of individuals infected over time;
\item $\frac{d\bar{R}}{dt}$ is the rate of change of the number of individuals recovered over time.
\end{itemize}

Because there is no canonical choice of $N$, we will transform the system \eqref{eq0} by dividing it by $N$ and considering $S(t)=\bar{S}(t)/N$, $I(t)=\bar{I}/N$ and $\hat{R}(t)=\bar{R}(t)/N$.  It is customary to choose $N=10^6$ for convenience but this is just an arbitrary choice.  For instance, analysis on smaller communities or cities involves less than $10^6$, however $10^6$ is a common choice because countries number their populations in multiples of $10^6$. With these notations we translate \eqref{eq0} into
\begin{equation}\label{eq1}
\begin{cases}
\frac{d S(t)}{dt}=-\beta S(t)I(t) \\
\frac{d I(t)}{dt}=-\beta S(t)I(t)-\gamma I(t) \\
\frac{d \hat{R}(t)}{dt}=\gamma I(t)
\end{cases}
\end{equation}
where $\beta=\bar{\beta}/N$ and $\gamma$ is the same as in \eqref{eq0}.

Notice that now we actually have that $S(t)+I(t)+\hat{R}(t)=S_0+I_0+\hat{R}_0=1$ for all $t\ge0$.  Since we are interested in the reverse problem, namely determining the parameters $\beta,\gamma$ from the observations, we put this as a formal mathematical result as follows.

\begin{proposition}\label{p:1}  Referring to the system \eqref{eq1}, if we know $I_0,S_0$ and the values $I(t_1), S(t_1)$ for some $t_1>0$, these determine uniquely the parameters $\beta$ and $\gamma$ of the system.
\end{proposition}

It is important to remark that one of the main assumption is that the parameters $\beta, \gamma$ do not change in time.

\subsubsection{The neural network}\label{s:2}

Our next goal is to get a rough round of estimates on the parameters $\beta,\gamma$ of the SIR model. For this we add the deaths to the recovered.  Taking into consideration the recommendations/restrictions that have been applied by the authorities, in almost all countries (school closure, the ban of public events, social distancing recommendation/constraint, self-isolation if experiencing symptoms, quarantine for people tested positive), we presume that these parameters \emph{are not} constant over time. We estimate the parameters based on two weeks of (cleaned) data.   

To train the neural network we first use the SIR model to simulate data.  We build a dataset based on the following procedure.  

\begin{enumerate}
\item Define a data set $\Delta$ to store the values generated below.
\item Take $A=\{i/50: i\in \{0,1,\dots,50\} \}$ and $B=\{ j/50: j \in \{1,2,\dots, 50\} \}$.  
\item Next we split the interval $(0,0.2)$ into $10$ equal subintervals: $C_j=(j/50,(j+1)/50)$ with $j=0,1,2,\dots,9$. 
In a similar way we define $D_k=(k/50,(k+1)/50)$ with $k=0,1,2,\dots,9$.

\item For each $\beta\in A$, $\gamma\in B$, $j\in\{0,1,2,\dots,9\}$ and $k\in\{0,1,2,\dots,9\}$
\item For each $l$ in $\{0,1,\dots, 200\}$ choose at random (uniformly)
\begin{enumerate}
    \item $I_0\in C_j$
    \item $R_0\in D_k$
    \item solve the SIR equation with parameters $\beta, \gamma$ and initial conditions $S_0,I_0,R_0$ for $t\in [0,14]$.  
    \item For each $t\in\{1,2,\dots, 14\}$ store in $\Delta$ the row $(\beta, \gamma, t,I_0,I(t), R_0,R(t))$.
\end{enumerate}

\end{enumerate}

Proposition~\ref{p:1} guarantees that if we know any values $I_0,I(t),R_0,R(t)$ we can uniquely determine the parameters $\beta, \gamma$.  We use the range of $t=1,2,\dots, 14$ to make the model more robust.  At the same time the two weeks period is also consistent with the average recovery time of an infected patient of Covid19 and also corresponds to the lawful period of reporting of data. 

We pick a sub sample (of size 80\%) from the rows of $\Delta$ and set  
\[
XTrain= \Delta(t,I_0,I(t),R_0,R(t))
\]
which are the columns of $\Delta$ corresponding to $t,I_0,I(t),R_0,R(t)$ and the output data is exactly the pair
\[
YTrain=\Delta(\beta,\gamma).
\]

The neural network we used is of the following form
\begin{enumerate}
 \item Dense 64, activation ReLU, with input dimension=4
 \item Dense 128, activation ReLU
 \item Dense 256, activation ReLU
 \item Dense 512, activation ReLU
 \item output $(\beta, \gamma)$ with optimizer Adam and loss MAE.
\end{enumerate}

\subsubsection{Daily estimates of the parameters}\label{s:ne}

\begin{figure}[H]
\begin{center}
    \includegraphics[width=.8\textwidth]{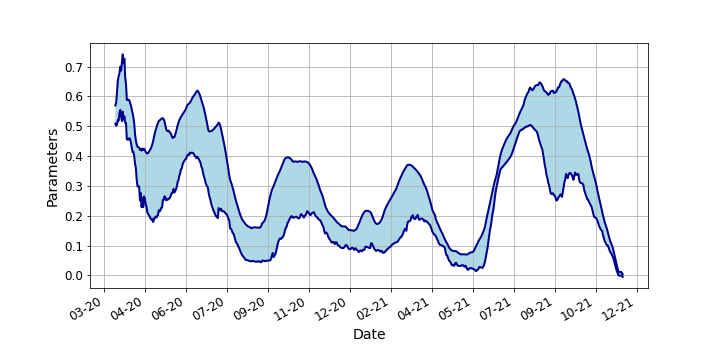}\hfill
    \includegraphics[width=.8\textwidth]{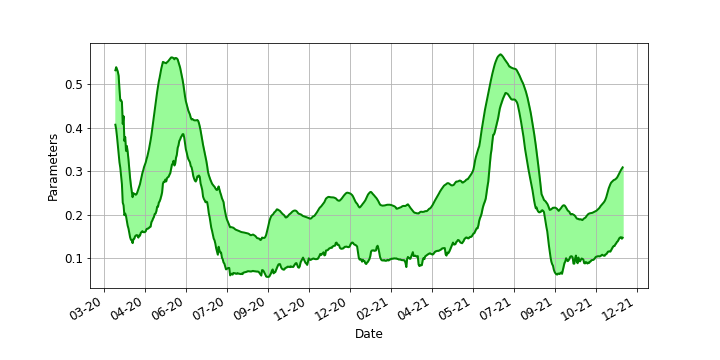}
\end{center}
 \caption{This is the plot of the daily estimated parameters $\beta$ and $\gamma$.  For each day $k$ we take 14 different estimates based on real data described above and we plotted the minimum and maximum value.}
 \label{f:3}
\end{figure}

To estimate the parameters from the real data, we proceed as follows.  For each day $k=0,1,2,\dots,T-14$ ($T$ is the data range) we use the neural network to predict the parameters $\beta_{k,t}, \gamma_{k,t}$ ($t=1,2,\dots, 14$) based on the real data $I_0=I_{real}(k)$, $I(t)=I_{real}(k+t)$, $R_0=R_{real}(k)$, $R(t)=R_{real}(k+t)$.  Thus for each day $k$ we determine $14$ estimates of the parameters which are plotted in Figure~\ref{f:3}. Here by $I_{real}$ and $R_{real}$ we refer to the cleaned real data.   

\begin{figure}[H]
  \centering
  \includegraphics[width=0.8\linewidth]{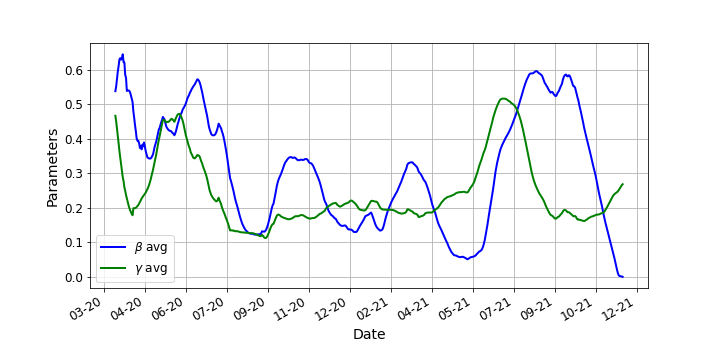}
 \caption{This is the plot of the average parameters $\beta$ and $\gamma$ by day.}
 \label{f:4}
\end{figure}

In Figure~\ref{f:4} we take the average of the parameters.  For each day $k$ we plot $\beta_k=\frac{1}{14}\sum_{t=1}^{14}\beta_{k,t}$ and similarly $\gamma_k=\frac{1}{14}\sum_{t=1}^{14}\gamma_{k,t}$.

We should also comment on the fact that the data that is available shows the number of individuals that have been tested positive, but it is very likely that the real number of people infected is in fact much higher, as there are also asymptomatic individuals, people that are not being tested although they present the specific symptoms, so they are not part of the official reports.

Thus the above predictions for the parameters constitutes a good starting point for an optimization procedure we describe now.  In order to reduce the effects of the above deficiencies, we consider another model which accounts for the number of deceased as a separate compartment. 

\subsection{Stage II - SIRD model and the parameter estimates}\label{s:stage2}

\subsubsection{The SIRD model}\label{s:3}

In the sequel we propose a model which refines the SIR model. The guiding line is that the number of deaths is the most reliable number we can account for, as the number of infected and recovered people could be largely unaccounted.

We have now four variables changing with time.  These are $S(t)$, $I(t)$, $R(t)$ and $D(t)$ where $R(t)$ is the proportion of recovered and alive people while the $D(t)$ is the proportion of deceased people.  We set the interaction as follows

\begin{equation}\label{sird}
\begin{cases}
\frac{d S(t)}{dt}=-\beta S(t) I(t)\\
\frac{dI(t)}{dt}=\beta S(t) I(t) -(\gamma_1+\gamma_2) I(t) \\
\frac{dR(t)}{dt}=\gamma_1 I(t)\\
\frac{dD(t)}{dt}=\gamma_2 I(t).
\end{cases}
\end{equation}

where:
\begin{itemize}
\item $\gamma_1$ = recovery rate
\item $\gamma_2$ = mortality rate
\end{itemize}

Notice that in this setup the removed population, from classical SIR model, $\hat{R}$, bifurcates into recovered ones, accounted by $R$ and the deceased ones accounted by $D$.  We can observe that $\hat{R}$ is the sum of the two factors $R+D$.  In this way we separate the dead people from the recovered ones which are mixed up in the classical SIR model and we are going to manipulate these equations and reduce the computations to a single equation involving only one of these quantities, the most reliable one, namely $D(t)$.  To do this we will write all the other quantities as functions of $D$ as follows:
\begin{equation*}
S=u(D), I=v(D), R=w(D).
\end{equation*}
The easiest to deal with is $R$ because from the last two equations we get
\begin{equation*}
\frac{dR}{dt}=\frac{\gamma_1}{\gamma_2}\frac{dD}{dt}
\end{equation*}
from which we deduce that $R(t)=\frac{\gamma_1}{\gamma_2}(D(t)-D_0)+R_0$.

Now, we deal with the function $u$ from $S(t)=u(D(t))$.  Dividing the first and the last we get
\begin{equation*}
u'(D)=-\frac{\beta}{\gamma_2}u(D)
\end{equation*}
which can be integrated and gives $S$ in terms of $D$ as
\begin{equation*}
S=S_0\exp\left( -\frac{\beta}{\gamma_2}(D-D_0) \right).
\end{equation*}

On the other hand this allows us to solve for $I=v(D)$.  First we notice that
\begin{equation*}
\frac{dS}{dt}+\frac{dI}{dt}=-(\gamma_1+\gamma_2)I=-\frac{\gamma_1+\gamma_2}{\gamma_2}\frac{dD}{dt}
\end{equation*}
from which we deduce that
\begin{equation*}
S+I+\frac{\gamma_1+\gamma_2}{\gamma_2}D=S_0+I_0+\frac{\gamma_1+\gamma_2}{\gamma_2}D_0.
\end{equation*}
therefore we obtain that (as functions of $D$)
\begin{equation}\label{e:D}
\frac{dD}{dt}=\gamma_2 I_0-(\gamma_1+\gamma_2)(D-D_0)+\gamma_2 S_0\left[1-\exp\left( -\frac{\beta}{\gamma_2}(D-D_0) \right)\right]
\end{equation}

This last implication works in the case the parameters $\beta,\gamma_1,\gamma_2$ are all assumed constant in time.  However, if they vary with time, then, the equation is a little bit different, the main equation becomes now
\begin{equation}\label{e:D2}
D'(t)=\gamma_2(t) I_0-\gamma_2(t)\int_0^t\left(\frac{\gamma_1(s)}{\gamma_2(s)}+1\right)D'(s)ds+\gamma_2(t) S_0\left[1-\exp\left( -\int_0^t\frac{\beta(s)}{\gamma_2(s)}D'(s)ds \right)\right]
\end{equation}

At this moment we can use the data on the death cases to estimate the parameters involved. As we pointed out already, the proportion of infected (or recovered) is grossly underestimated since there are probably more infected people than the reported cases tested.  

From the technical standpoint, equation \eqref{e:D2} is not easy to handle and we will use equation \eqref{e:D} instead together with the implicit assumption that the parameters are constant for short periods of time.  More precisely, in our approach we take the time interval on which we assume the parameters constant to be two weeks, which is in accordance with the lawful time of reporting and also with the dynamic of the time to recovery. In other words, we fit the number of deceased on pieces of two weeks where we assume that the parameters do not change.  

\subsubsection{Parameter estimates}\label{s:sirdest}

The estimation of the parameters is done using a grid search based on the values already found with the SIR model estimates.  The main reason is that we now use the previously found set of parameters as the starting point of the grid search.  We do this dynamically, starting with any given day $k$ and use the next $14$ days forward to search for the set of parameters $I_k,R_k,\beta_k,\gamma_{1,k}, \gamma_{2,k}$ to find the ones which best predict the number of deaths. 

We detail here the main steps.  

\begin{enumerate}
\item For each day $k$ we take the next $14$ days of data. 
\begin{enumerate}
    \item Compute $\bar{\beta}_k=\frac{1}{14}\sum_{i=k}^{k+14}\beta_i$, which is the average of the parameters $\beta_k$ discussed in the previous paragraph.  Similarly we consider $\bar{\gamma}_k=\frac{1}{14}\sum_{i=k}^{k+14}\gamma_i$.  
    \item Take 
    \begin{itemize}
        \item $E=\{(\frac{7}{10}+\frac{3*i}{100})\times \bar{\beta}_k: i\in \{0,1,\dots,19\} \}$,  
        \item $F=\{(\frac{7}{10}+\frac{3*i}{100})\times \bar{\gamma}_k: i\in \{0,1,\dots,19\} \}$, 
        \item $G=\{(\frac{7}{10^4}+\frac{3*i}{10^5})\times \bar{\gamma}_k: i\in \{0,1,\dots,199\} \}$
        \end{itemize}
    \item Now we solve by grid search the argmin problem 
    \begin{equation}\label{e:min1}
    \begin{aligned}
    & l^*_k,m^*_k,n^*_k=\underset{l\in E, m\in F, n\in G}{\mathrm{argmin}}
    & & \sum_{t=k}^{t=k+14} \bigg(D(t)-Data(t)\bigg)^2 \\
    \end{aligned}
    \end{equation}
    where $D(t)$ is the solution of \eqref{e:D} and $Data(t)$ is the reported number of dead individuals.
\end{enumerate}
\end{enumerate}

\begin{figure}[H]
  \centering
  \includegraphics[width=0.8\linewidth]{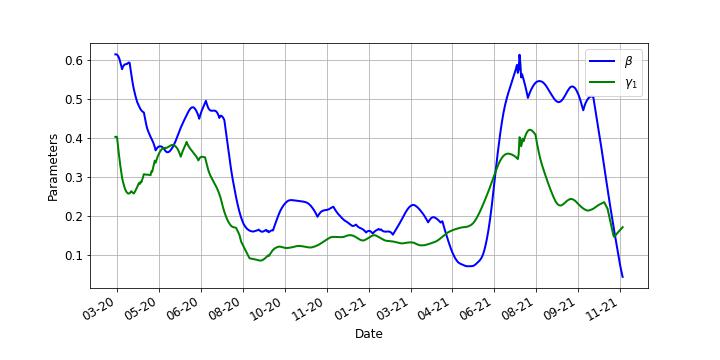}
 \caption{This is the plot of the parameters $\beta$, $\gamma_1$ by day.}
 \label{f:5}
\end{figure}

\begin{figure}[H]
  \centering
  \includegraphics[width=0.8\linewidth]{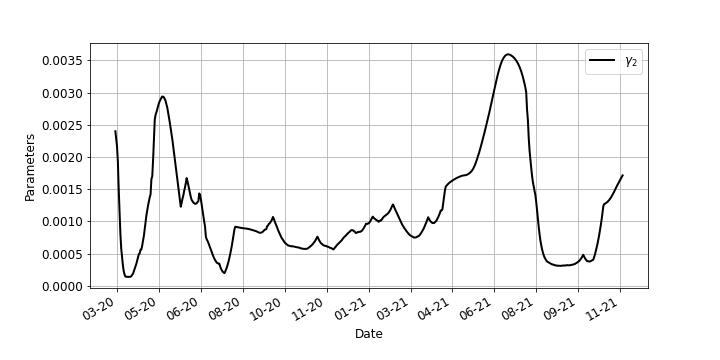}
 \caption{This is the plot of the parameter $\gamma_2$ by day.}
 \label{f:6}
\end{figure}

Figures \ref{f:5} and \ref{f:6} illustrate the evolution of the optimized parameters $\beta,\gamma_1,\gamma_2$ using \eqref{e:min1}.

\subsection{Stage III - Regimes and regime switch}\label{s:stage3}

We define first a \emph{turning point} or a \emph{regime switch} as a time where the parameters attain a local extreme (maximum or minimum).   \emph{A regime is a period between two consecutive turning points}.    

To model the dynamic inside a regime, we use the linear regression in order to make prediction for the future periods.  

Given times $T=\{t_1,t_2,\dots, t_k\}$ and positions $X=\{x_1,\dots, x_k \}$ we define the regression line obtained using the least square optimization.  Precisely, we set:  
\begin{equation}\label{e:reg}
\begin{split}
\sigma(t,T,X)&=a+bt \text{ where } \\ 
a&=\frac{(\sum_{i=1}^k t_i^2)(\sum_{i=1}^k x_i)-(\sum_{i=1}^k t_i)(\sum_{i=1}^k t_i x_i)}{k(\sum_{i=1}^k t_i^2)-(\sum_{i=1}^k t_i)^2} \\ 
b&=\frac{k(\sum_{i=1}^k t_i x_i)-(\sum_{i=1}^k t_i) (\sum_{i=1}^k x_i))}{k(\sum_{i=1}^k t_i^2)-(\sum_{i=1}^k t_i)^2}.  
\end{split}
\end{equation}

We talked about the existence of different regimes in the spreading of Covid19 because of the measures that have been taken, which had a significant impact on the evolution of the infection rate.  We consider now a regime starting at time $p_1$, with a set of parameters, ending at time $p_3$ and having the intermediary time $p_2$. We adapt the \textbf{SIRD model} as follows
\begin{equation}\label{sird_regimes}
\begin{cases}
\frac{d S(t)}{dt}=-\sigma_\beta(t)\cdot  S(t)\cdot I(t)\\
\frac{dI(t)}{dt}=\sigma_\beta(t)\cdot S(t)\cdot I(t)-(\sigma_{\gamma_1}(t)+\sigma_{\gamma_2}(t))\cdot I(t)\\
\frac{dR(t)}{dt}=\sigma_{\gamma_1}(t)\cdot I(t)\\
\frac{dD(t)}{dt}=\sigma_{\gamma_2}(t)\cdot I(t).
\end{cases}
\end{equation}
where $\sigma_{\beta},\sigma_{\gamma_1}, \sigma_{\sigma_2}$ represent the regression functions based on a number of consecutive days and the parameters $\beta,\gamma_1,\gamma_2$ evaluated at these days obtained from the optimization scheme \eqref{e:min1}. More precisely, we take for a fixed day $k\ge7$, the previous $7$ days and we use the regression line based on these days and the values of the estimated parameters. The effect of using the regression line has a smoothing effect on the parameters. As we plainly see, the parameters are not constant in time, not even for relatively short periods of time (say for instance 30 days).  This new model is robust to the fluctuation of the parameters.         

\begin{figure}[H]
 \centering
\includegraphics[width=.8\textwidth]{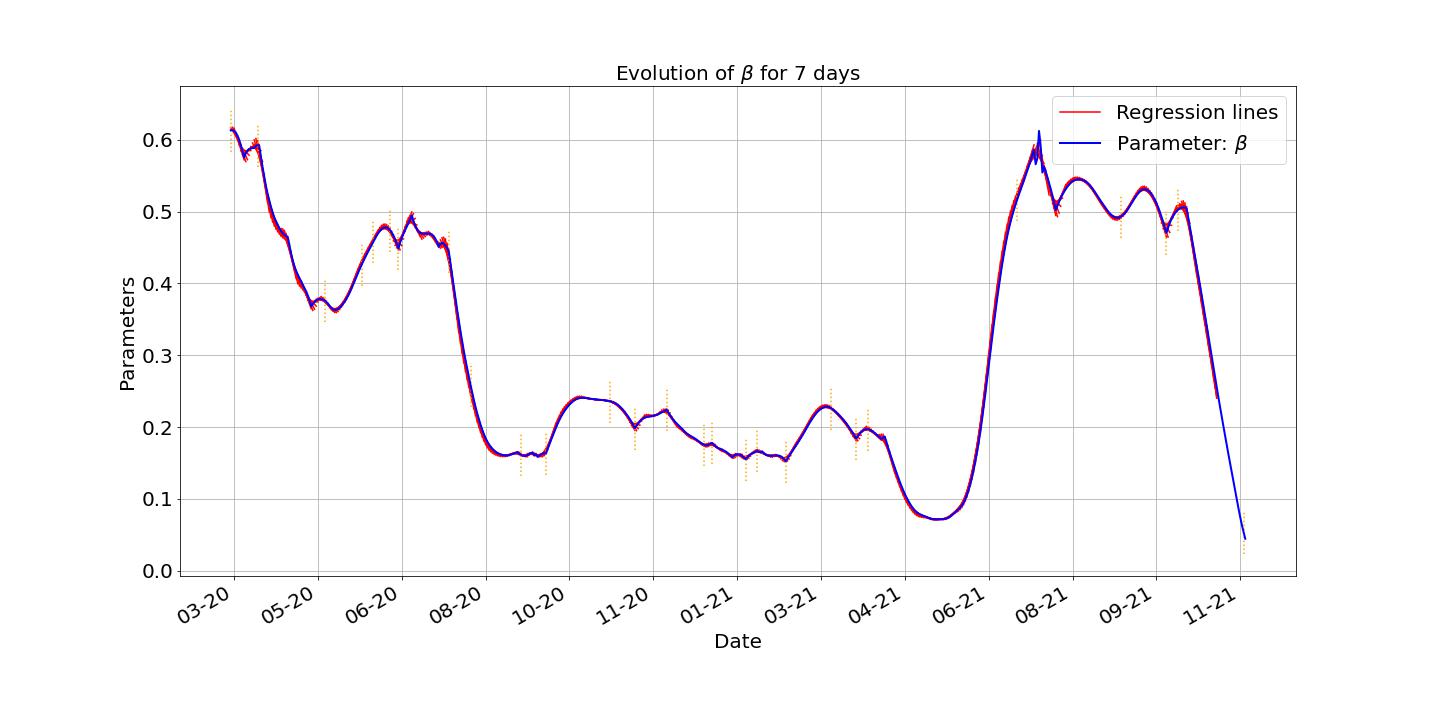}\hfill
\includegraphics[width=.8\textwidth]{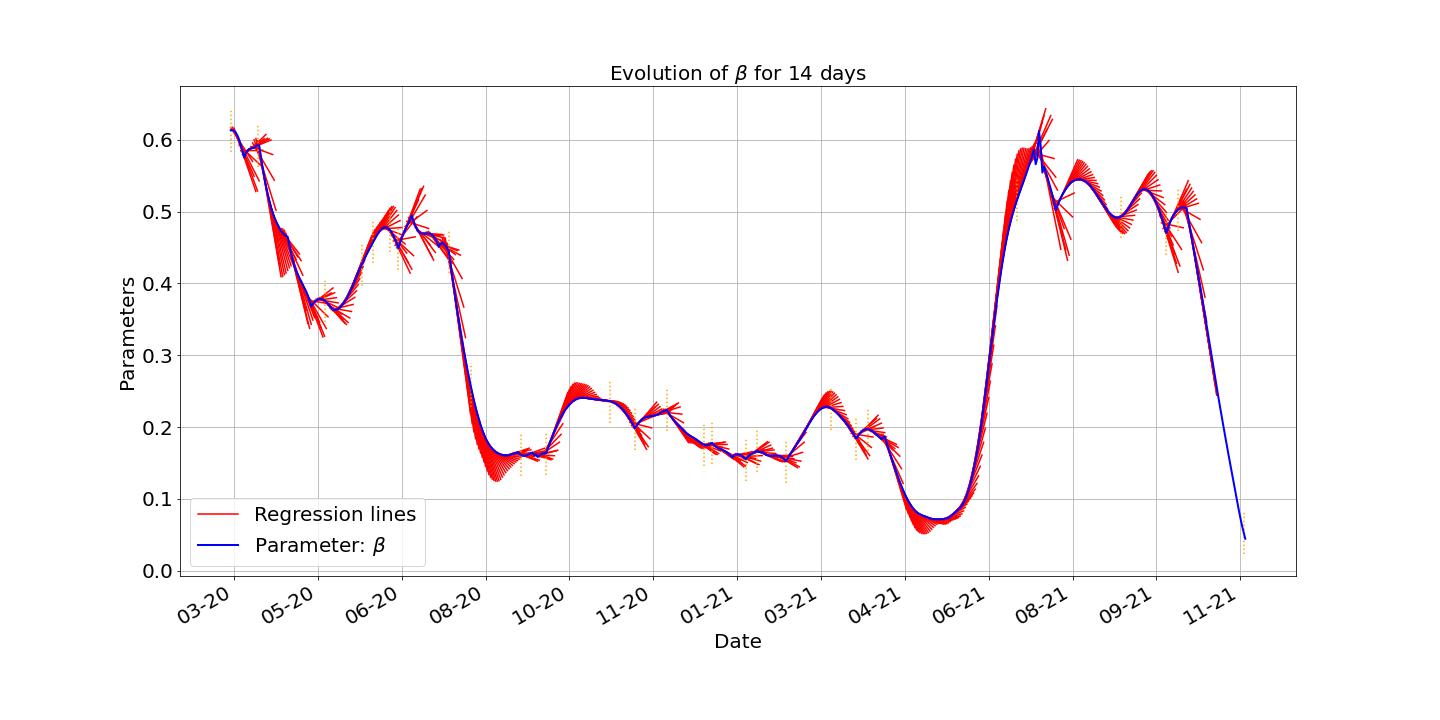}
\caption{The first plot shows the estimated parameter $\beta$ together with the regression lines started at each day, using the last $7$ days data and plotted for $7$ days.  In the second one, we plotted the regression line for 14 days, based on 7 days from the data.  }
\label{e:fig8.1}
\end{figure}

\begin{figure}[H]
 \centering
\includegraphics[width=.8\textwidth]{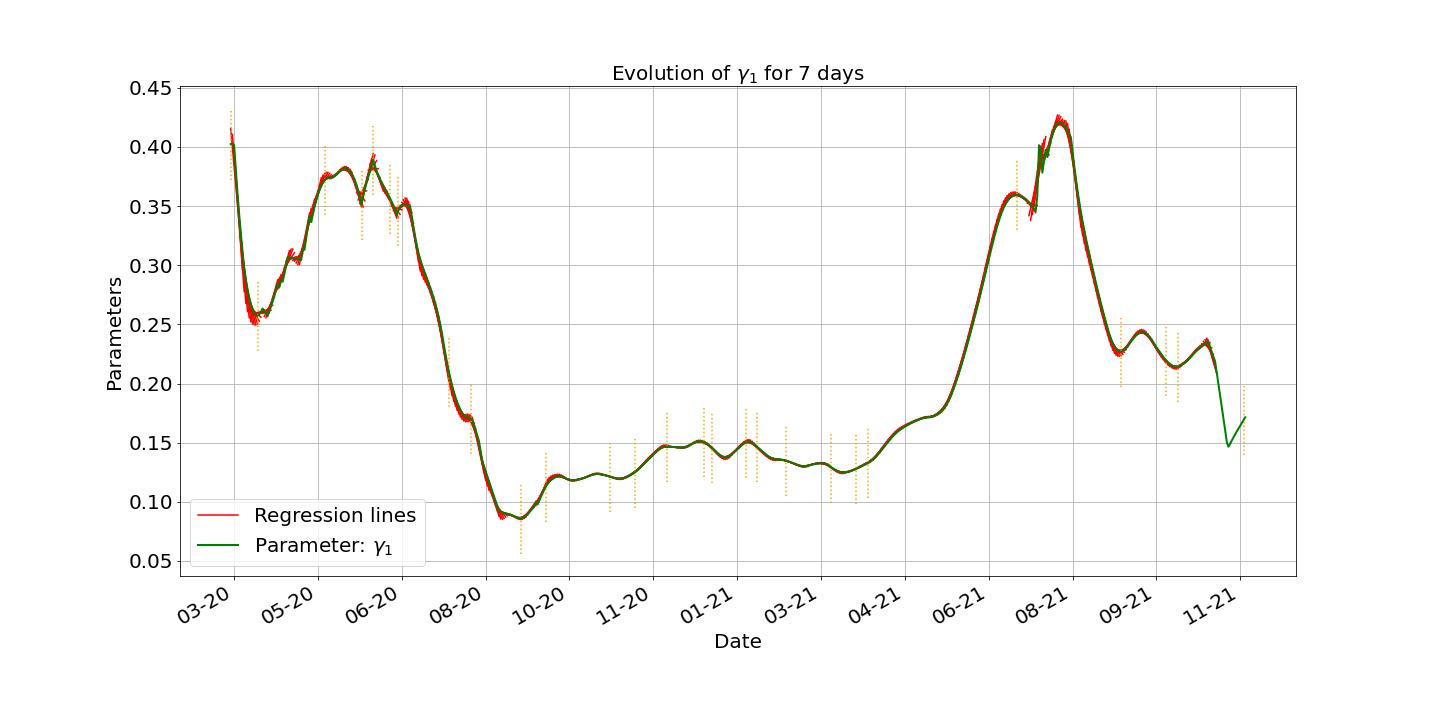}\hfill
\includegraphics[width=.8\textwidth]{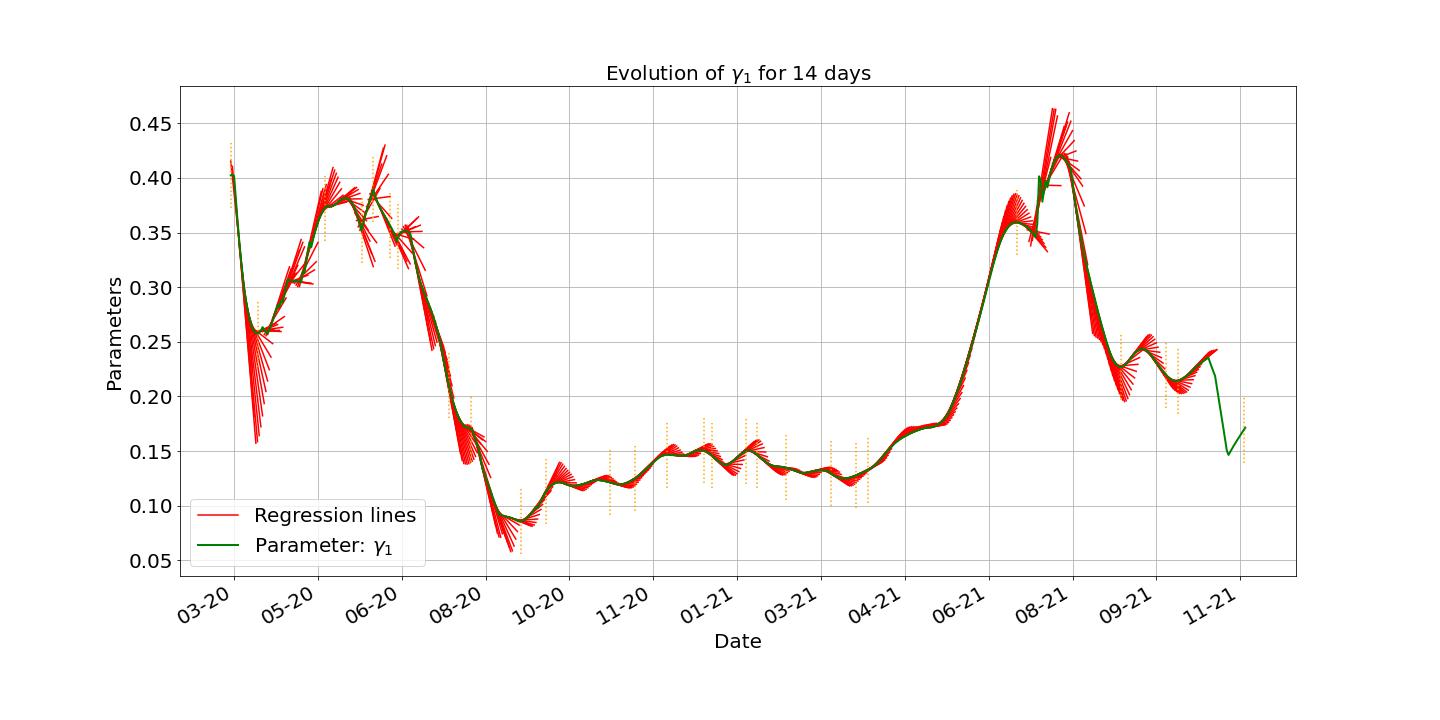}
\caption{The first plot shows the estimated parameter $\gamma_1$ together with the regression lines started at each day, using the last $7$ days data and plotted for $7$ days.  In the second one, we plotted the regression line for 14 days, based on 7 days from the data.  }
\label{e:fig8.2}
\end{figure}

\begin{figure}[H]
 \centering
\includegraphics[width=.8\textwidth]{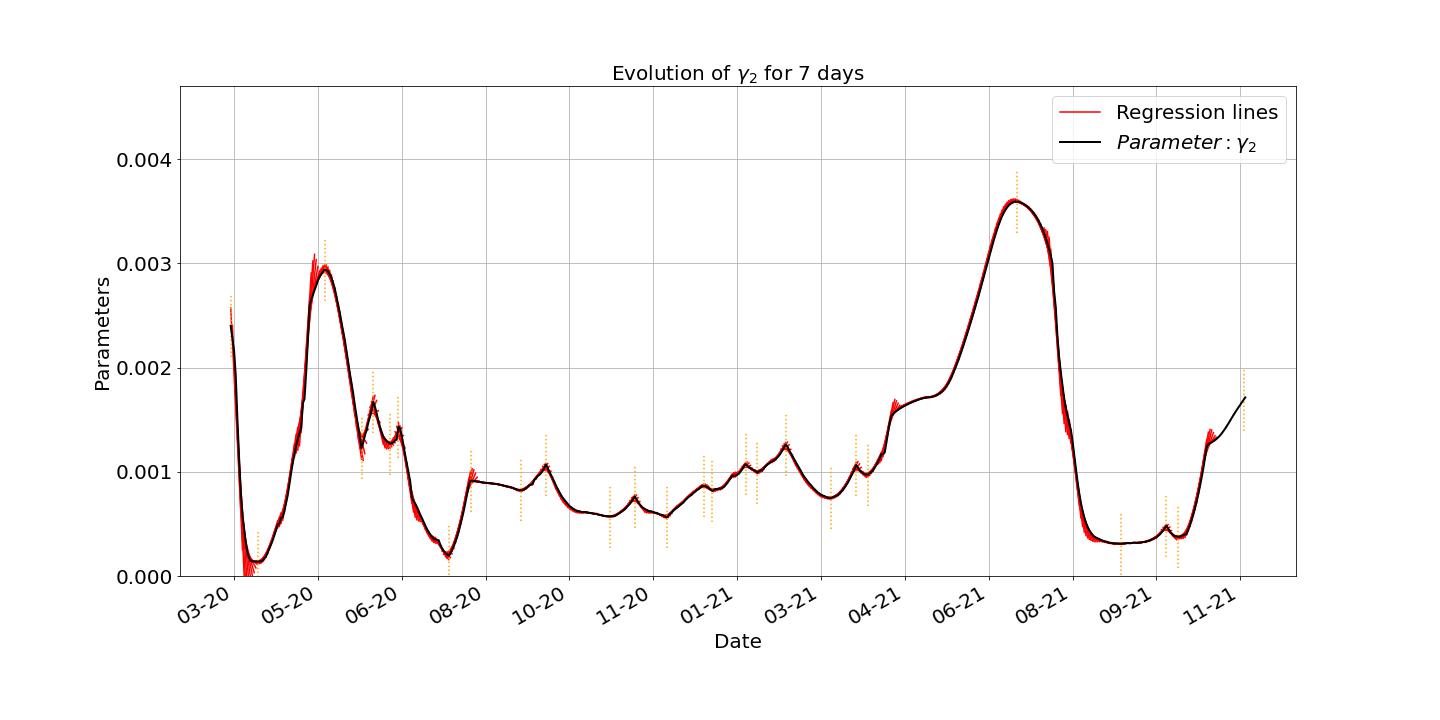}\hfill
\includegraphics[width=.8\textwidth]{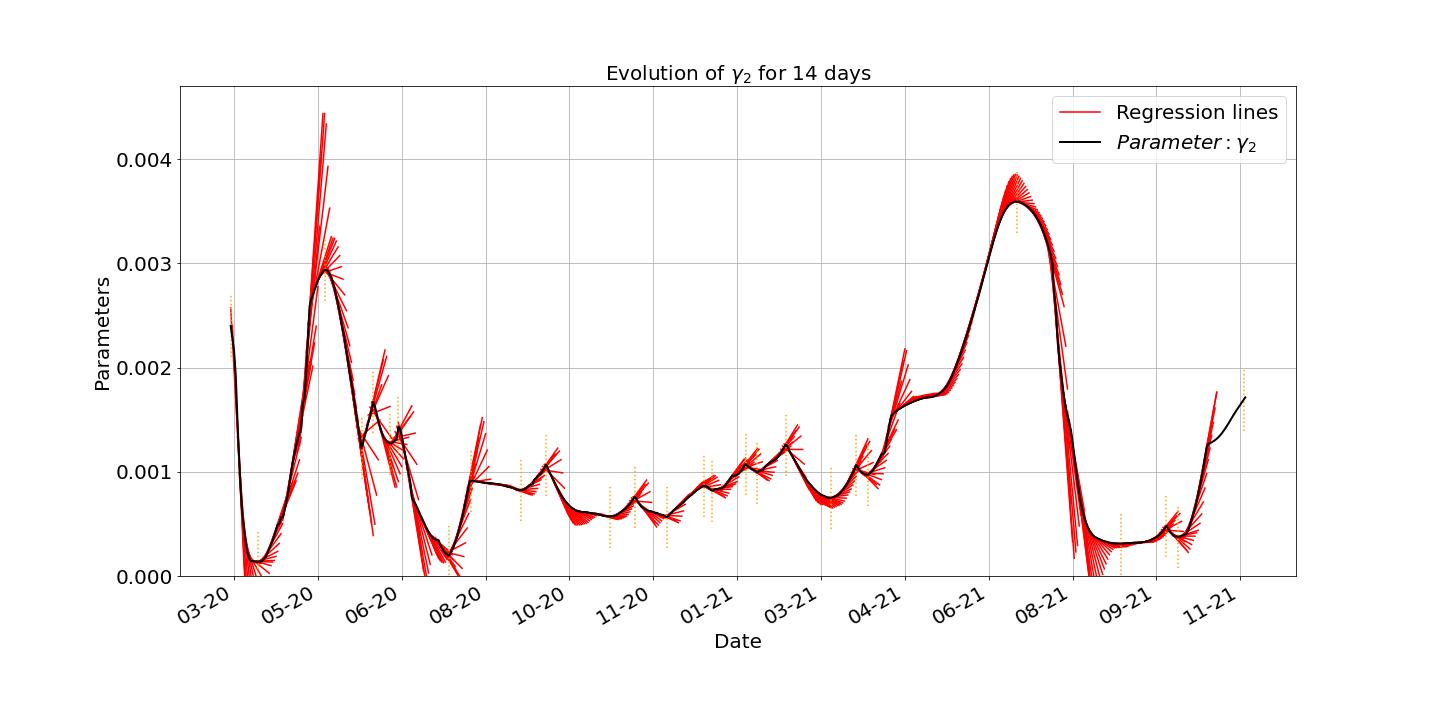}
\caption{The first plot shows the estimated parameter $\gamma_2$ together with the regression lines started at each day, using the last $7$ days data and plotted for $7$ days.  In the second one, we plotted the regression line for 14 days, based on 7 days from the data.  }
\label{e:fig8.3}
\end{figure}

Below, in Figures~\ref{e:fig8.1}, \ref{e:fig8.2}, \ref{e:fig8.3} we have the evolution of the estimates obtained in \eqref{e:min1} together with the regression lines constructed in terms of $7$ days.  It is important to remark that the predictions lose their power around the turning points. In each of the cases we have a good indication of the accuracy with which the regression predicts the behavior of the parameters.  With vertical dotted line we mark the points where we have a local extreme (minimum or maximum) in the evolution of $\gamma_2$.  The turning days are those days $k$ for which the value of $\gamma_2(k)$ is an extreme value for the period $[k-7,k+7]$ days centered at that given $k$ day.   

\section{Predictions}\label{s:pred}  

Based on the method outlined above we predict the proportion of deaths into the future.  The idea is to use now the system \eqref{sird_regimes}.  More precisely, for any given day $k$ we use
the following recipe:  
\begin{enumerate}
\item Use $7$ previous days of the parameters produced by the estimation procedure from \eqref{e:min1} to get the regression functions $\sigma_{\beta},\sigma_{\gamma_1},\sigma_{\gamma_2}$.  
\item Based on these data, we solve the system \eqref{sird_regimes} for the next 14 or 21 days to see the fit with the real data of deceased.  
\end{enumerate}

The results are illustrated below in Figure \ref{e:fig9}.  

\begin{figure*}[ht!]
\begin{center}
\includegraphics[width=.8\textwidth]{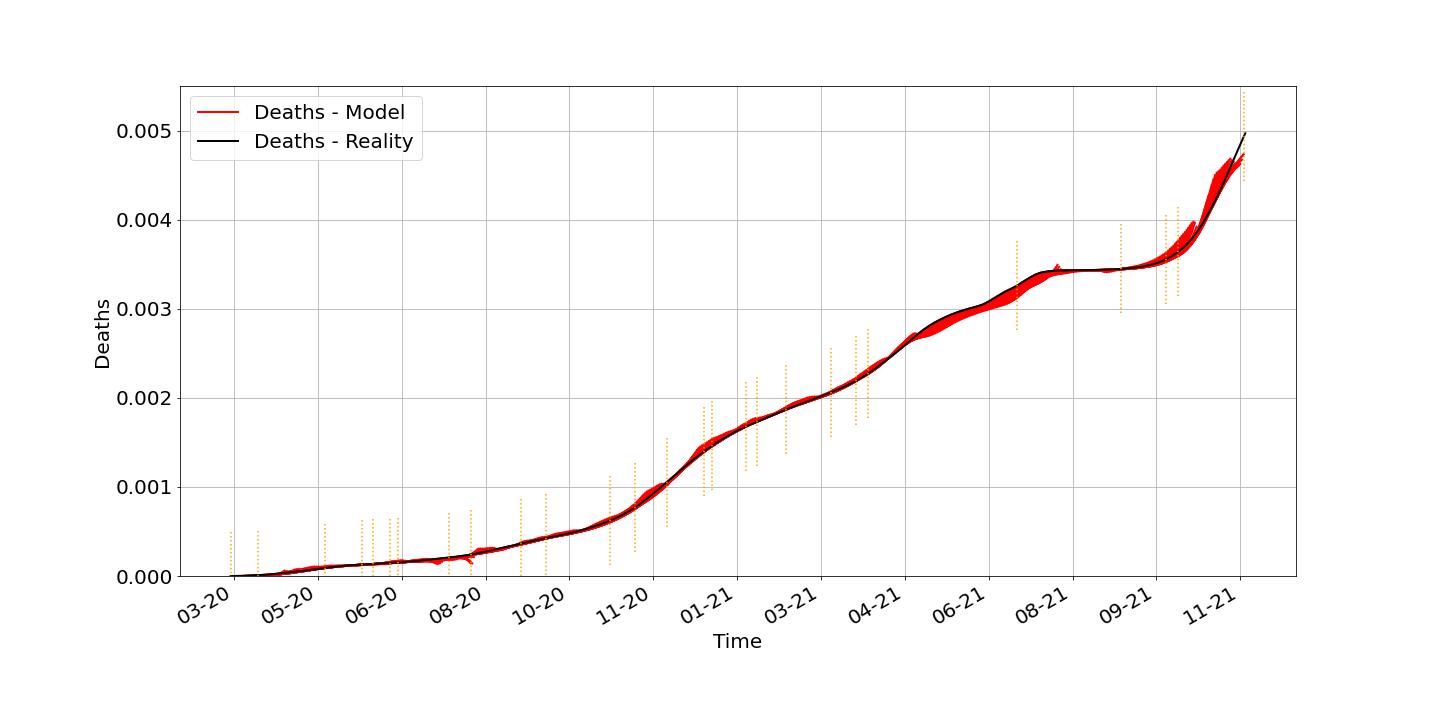}\hfill
\includegraphics[width=.8\textwidth]{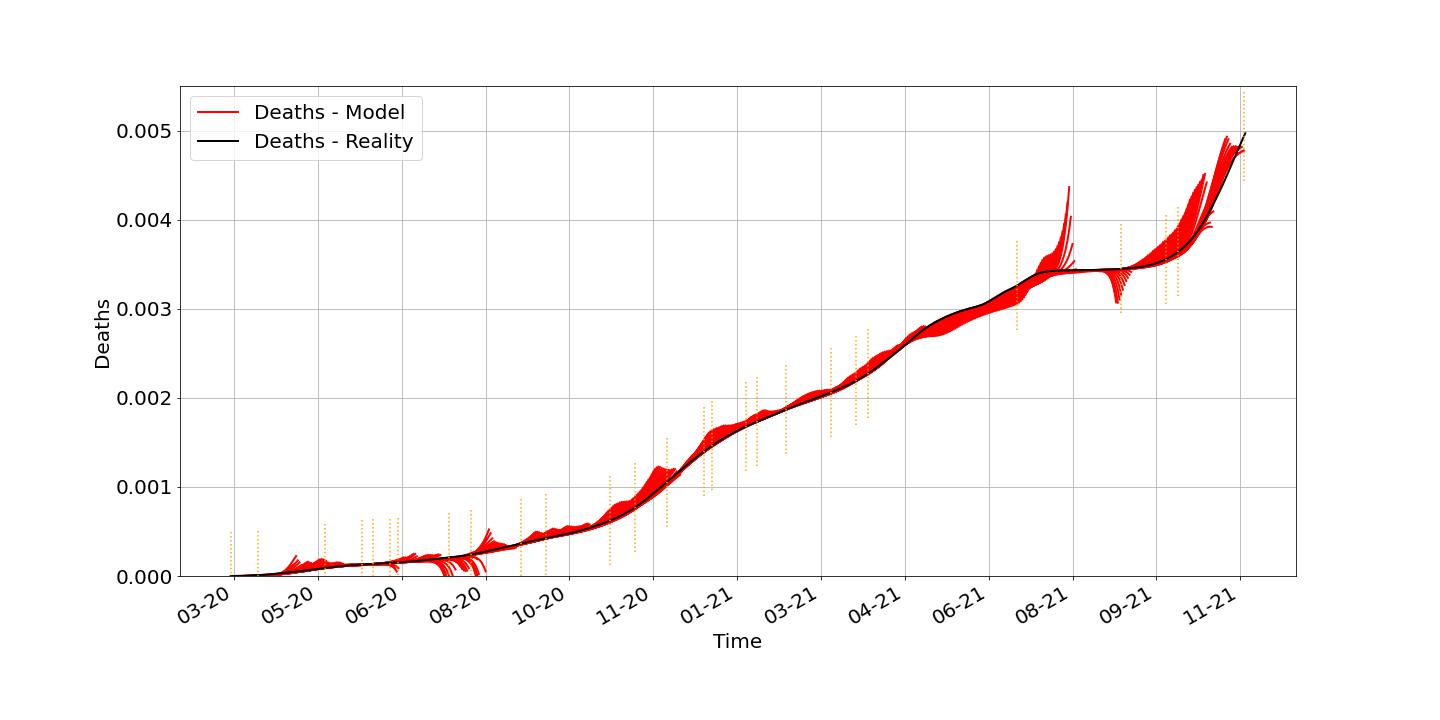}\hfill
\end{center}
\caption{This figure shows the predictions (in red) starting at any given day $k$ in which we estimate using a standard regression for the past $7$ days and use the system \eqref{sird_regimes} to predict for 14 days (first one) and 21 days (second one) in the future.  We find important to mention that we achieve better prediction for shorter period (14 days versus 21 days). In both cases, we observe that the prediction is better between the turning points. }
\label{e:fig9}
\end{figure*}

\section{Conclusions}\label{s:6}

One of the characteristic of the Covid19 pandemic is that there were continuous changes of the conditions due to many factors.  For instance, we can mention the quarantine periods, relaxation, mask enforcement, school openings and closings, vaccination campaign, major events, new variants of the virus, time of the year.  If we take any parametric model, it is natural to assume that the parameters of the model change over time.   

On the other hand, the reported numbers of infected and recovered during the pandemic is not reliable. Because of this reason we base our estimates on the reported number of deceased people which is, in our opinion, more realistic.  However, discarding the data of infected and recovered entirely is not reasonable.  

In this light, we proposed a three layers approach to the analysis of the pandemic of Covid19.

The first layer consists in using the SIR model to estimate the first round of its parameters, $\beta$ and $\gamma$ assumed constant for periods of up to $14$ days.  Technically, we do this using a neural network trained on data simulated with the SIR model.  Using this neural network and real data for any period of 14 days we generate a pair $(\beta,\gamma)$ which best describes the period from the point of view of the SIR model.    

The second layer is based on a modification of the SIR model to account for the dead individuals.  We call this SIRD and to our knowledge it is not used in this framework in the literature (though it is also recently used in cite{}). The SIRD model is depending on three parameters $\beta,\gamma_1,\gamma_2$.  Based on the result of the previous layer, we use an optimization problem to fit the number of reported deaths. We do this procedure assuming the parameters are constant on pieces of $14$ days.  The result provides, for any given day, a prediction of the parameters of the model using the previous 14 days.  

The last layer consists in using the parameters to determine the turning points, the extreme points of the parameters (more precisely we do this for $\gamma_2$).  We define a regime as the time period between two turning points. This perspective is consistent with our assumption that the parameters are not constant for long periods of time. 

This last layer, combined with an adaptation of the SIRD model to account for the change in parameters, leads to a way to predict the evolution of the pandemic, at least for short periods of time.  

We believe that this proposed methodology is a general one and can be extended to the analysis of spread of Covid19 in any country provided that we have relevant data.  This methodology can also be extended to many  other diseases which can be modeled by SIR and SIRD.

\section{Declaration}

\subsection{Ethics approval} We did not use any confidential data for the analysis in this paper and we do not have any ethical issues in this paper.

\subsection{Consent for publication}

We did not use any data which could possibly reveal any personal data of any patient.

\subsection{Availability of data and material} We used the public data from \href{https://datahub.io/core/covid-19/r/3.html}{here}.

\subsection{ Competing interests } The authors declare that they have no competing interests.

\subsection{ Funding } There are no funding sources for this paper.

\subsection{Authors' contributions}  All authors contributed equally to this paper.

\subsection{Acknowledgement}  The last author would like to thank Iulian Cimpean, Lucian Beznea and Mihai N. Pascu for useful discussions about this paper.  Many thanks to the reviewers of this paper for the comments and directions which led to a much better version of this paper.  
\providecommand{\bysame}{\leavevmode\hbox to3em{\hrulefill}\thinspace}
\providecommand{\MR}{\relax\ifhmode\unskip\space\fi MR }
\providecommand{\MRhref}[2]{%
  \href{http://www.ams.org/mathscinet-getitem?mr=#1}{#2}
}
\providecommand{\href}[2]{#2}

\newcommand{\etalchar}[1]{$^{#1}$}
\providecommand{\bysame}{\leavevmode\hbox to3em{\hrulefill}\thinspace}
\providecommand{\MR}{\relax\ifhmode\unskip\space\fi MR }
\providecommand{\MRhref}[2]{%
  \href{http://www.ams.org/mathscinet-getitem?mr=#1}{#2}
}
\providecommand{\href}[2]{#2}

\section{Appendix}\label{t:2}

The goal of this section is to provide the proof of Proposition~\ref{p:1}. Recall the system \eqref{eq1} given by

\begin{equation}
\begin{cases}
\frac{d S}{dt}=-\beta SI \\
\frac{d I}{dt}=-\beta SI-\gamma I \\
\frac{d R}{dt}=\gamma I
\end{cases}
\end{equation}

\begin{proposition} Referring to the system \eqref{eq1}, if we know $I_0,S_0$ and the values $I(t_1), S(t_1)$ for some $t_1>0$, these determine uniquely the parameters $\beta$ and $\gamma$ of the system.
\end{proposition}

Notice there the main assumption, that the parameters $\beta, \gamma$ do not change in time.

\begin{proof}
The first step is to notice that by assumption, $\beta, \gamma$ constants in time yields in the first place that
\begin{equation*}
\frac{I'(t)}{S'(t)}=-\frac{\beta S(t)I(t)-\gamma I(t)}{\beta S(t)I(t)}=-1+\frac{\gamma}{\beta S(t)}.
\end{equation*}
which in turn gives that
\begin{equation*}
I'(t)=-S'(t)+\frac{\gamma}{\beta}\frac{S'(t)}{S(t)}
\end{equation*}
and finally integrating this shows that (here we denote $\rho=\gamma/\beta$)
\begin{equation*}
I(t)+S(t)-\rho \log(S(t)) \text{ constant in }t.
\end{equation*}
In particular, this means that
\begin{equation}\label{e0:1}
I(t)+S(t)-\rho\log(S(t))=I_0+S_0-\rho \log(S_0).
\end{equation}
Typically the initial value of $S_0$ is close to $1$ and $I_0$ is relatively small. In particular, if we assume that the epidemic ends somewhere then we definitely have $I(t)=0$ and thus $S(t)$ solves the equation
\begin{equation}
S-\rho \log(S)=I_0+S_0-\rho \log(S_0).\label{eq2}
\end{equation}
In particular if we assume that $I(t_\infty)=0$ and $S(t)$ converges as $t\to t_\infty$, then we get in the limit that $S(t_\infty)$ solves \eqref{eq2}.  One consequence of this argument is that for all time $0\le t\le t_{\infty}$, we have  that $S(t)-\rho\log(S(t))\le \alpha:= I_0+S_0-\rho \log(S_0)$.

Another important consequence of this model is that if we assume $S_0$ and $I_0$ fixed (obviously $R_0$ will also be determined) but, for a given time $t=t_1>0$, knowing $S(t_1)$ and $I(t_1)$ (therefore $R(t_1)$ as well), we can determine uniquely the parameters $\beta$ and $\gamma$.  Indeed this is clearly seen from \eqref{e0:1} which gives
\[
 \rho=\frac{I_0+S_0-I(t_1)-S(t_1)}{\log(S_0)-\log(S(t_1))}.
\]

On the other hand, from \eqref{e0:1} in the first line of \eqref{eq1}, then we obtain that
\begin{equation}\label{e0:S}
 \frac{dS}{dt}=-\beta S(I_0+S_0-\rho\log(S_0) - S +\rho \log(S))
\end{equation}
The problem is that we can not integrate explicitly this to obtain an analytic expression for $S(t)$. However, what we can still show is that by knowing $I_0,S_0,I(t_1),S(t_1)$ we can determine the parameter $\beta$.  As we already pointed out, we know how to determine $\rho=\gamma/\beta$, thus we can rewrite \eqref{e0:S} in the form
\begin{equation}\label{e0:S2}
 \frac{S'}{S(I_0+S_0-\rho\log(S_0)-S+\rho\log(S))}=-\beta.
\end{equation}
Now, for $\alpha:=I_0+S_0-\rho\log(S_0)>0$ and $\rho>0$ we define for $x$ such that $\alpha>x-\rho \log(x)$,
\[
 \Phi(x)=\int_x^1\frac{ds}{s(\alpha-s+\rho \log(s))}
\]
and notice that using this function, integrating \eqref{e0:S2}, we arrive at
\[
 \Phi(S(t_1))-\Phi(S(0))=\beta t_1
\]
from which it is clear that $\beta$ is completely determined by $S(t_1),S_0,I_0$.  Knowing $\beta$ and $\rho$, we can immediately solve for $\gamma=\rho\beta$, thus all parameters are determined.

\end{proof}

\end{document}